\documentclass[
onecolumn,
amsmath,amssymb,
nofootinbib,
showkeys,
prd,
aps,
10pt,
longbibliography,preprintnumbers
]{revtex4-2}
\pdfoutput=1
\usepackage{hyperref}
\usepackage{xcolor,graphicx}
\usepackage{bm}

\begin{document}
	
\title{Black hole thermodynamics probes the equivalence principle}

\author{Ana Alonso-Serrano}
\email{ana.alonso.serrano@aei.mpg.de}
\affiliation{Institut für Physik, Humboldt-Universität zu Berlin, Zum Großen Windkanal 6, 12489 Berlin, Germany}
\affiliation{Max-Planck-Institut f\"ur Gravitationsphysik (Albert-Einstein-Institut), \\Am M\"{u}hlenberg 1, 14476 Potsdam, Germany}

\author{Luis J. Garay}
\email{luisj.garay@ucm.es}
\affiliation{Departamento de F\'{i}sica Te\'{o}rica and IPARCOS, Universidad Complutense de Madrid, 28040 Madrid, Spain}

\author{Marek Li\v{s}ka,\footnote{Corresponding author}}
\email{liskama4@stp.dias.ie}
\affiliation{School of Theoretical Physics, Dublin Institute for Advanced Studies, 10 Burlington Road, Dublin 4, Ireland}

\begin{abstract}
The equivalence principle for test gravitational physics strongly constrains dynamics of spacetime, providing a powerful criterion for selecting candidate theories of gravity. However, checking its validity for a particular theory is often a very difficult task. We devise here a simple theoretical criterion for identifying equivalence principle violations in black hole thermodynamics. Employing this criterion, we prove that Lanczos-Lovelock gravity violates the strong equivalence principle, leaving general relativity as the only local, diffeomorphism-invariant theory compatible with it. However, we also show that certain nonlocal expressions for black hole entropy appear to obey the strong equivalence principle.
\end{abstract}

\maketitle

\vspace{0.1cm}

Essay written for the Gravity Research Foundation 2023 Awards for Essays on Gravitation

\newpage

The equivalence principle amounts to the seemingly simple statement that the, suitably defined, test physics does not locally depend on the spacetime curvature. Despite its apparent simplicity, it has played a crucial role in the development of the immensely successful general relativity~\cite{Einstein:1911}. Today, the equivalence principle again offers a simple and robust selection criterion which can guide the development of the myriad alternatives to general relativity and, ultimately, of quantum gravity~\cite{Thorne:1973}. Working in this direction, numerous authors have introduced a hierarchy of variously stringent versions of the equivalence principle, devising their precise, operational statements~\cite{Ehlers:1972,Ehlers:1973,Thorne:1973,Geroch:1975,Casola:2014,Casola:2015,Fletcher:2022a,Fletcher:2022b,Wheeler:2024}. And, crucially, they have shown how the various equivalence principles allow us to introduce structures in the spacetime~\cite{Ehlers:1972,Wheeler:2024}, and how they in turn constrain the dynamics of both the matter fields~\cite{Ehlers:1973,Geroch:1975,Fletcher:2022b} and the spacetime itself~\cite{Thorne:1973,Casola:2014,Casola:2015}. In this way, the equivalence principles, while often considered to be nebulous and impractical~\cite{Synge:1960}, have in fact matured into a useful tool for studying gravity.

Herein, we contribute to this exploration by noting the close connection between the equivalence principles and thermodynamics. For instance, the thermodynamic description of a gas in a gravitational field relies on the equivalence of gravitational and inertial masses of the gas particles. Also, a uniformly accelerating observer in a curved spacetime sees a thermal bath of particles at the Unruh temperature, provided that the local inertial vacuum is well approximated by the flat spacetime one~\cite{Fulling:1973,Bisognano:1976,Unruh:1976,Barbado:2012}. And Tolman law for the equilibrium temperature in a curved spacetime, being encoded in the gravitational redshift, follows directly from the equivalence principle~\cite{Santiago:2018a,Santiago:2018b}.

The aforementioned examples concern the gravitational kinematics, i.e., behavior of matter in a fixed curved background. We instead focus on a thermodynamic quantity directly determined by the gravitational dynamics: the entropy of black holes~\cite{Bekenstein:1973,Bardeen:1973,Wald:1993}. We argue that measurement of changes in black hole entropy provides a conceptually simple check for violations of the strong equivalence principle. This most stringent statement of the equivalence principle asserts that all fundamental test physics (including gravitational phenomena) is locally unaffected by the background gravitational field~\cite{Casola:2015}. By being locally unaffected we mean that for every local experiment performed in a curved spacetime, one can find a suitable reference frame in the flat spacetime in which the experiment yields the same result. The test physics is any process whose backreaction on the spacetime can be safely neglected. The word ``fundamental'' presents a thornier problem~\cite{Casola:2015,Fletcher:2022b}. In practice, it means that we exclude processes governed by averaged, effective equations, such as dynamics of composite bodies. These phenomena generically depend, even locally, on the tidal gravitational forces and cannot be reconciled with the equivalence principle.

On the level of classical gravitational dynamics, at which we choose to stay for the time being, black hole thermodynamics is certainly fundamental in this sense. Its first law is encoded in the symplectic structure of the theory and requires no averaging procedure~\cite{Wald:1993,Wald:1994}. The Hawking effect providing the notion of black hole temperature then represents a direct consequence of quantum field theory in a curved spacetime~\cite{Hawking:1975,Visser:2003}.

In the following, we first describe a thought experiment that allows us to measure changes in black hole entropy and use them to probe the strong equivalence principle. Then, we consider the outcomes of this experiment for some gravitational theories of interest.

\section*{Setup of the thought experiment}

Let us suppose that we locate (or, in a far bolder scenario, manufacture) a black hole and use it as a test particle for the strong equivalence principle. Around this black hole, we consider a spatial ball of geodesic radius $l\gg r_+$, with $r_+$ being the radius of the outermost apparent horizon\footnote{The concept of an event horizon, being a global characteristic of the spacetime, does not apply to our local test black hole.}. Outside this ball, we can safely neglect the black hole's gravitational field as it scales with $r_+/l\ll 1$. We require that the ball is in vacuum. Otherwise, the black hole's gravitational pull on the background matter would be strong, possibly leading to substantial changes of the spacetime at long time scales. Moreover, to be able to meaningfully separate the black hole's gravitational field from that of the background spacetime, we require that the ball's radius $l$ is much smaller than the local curvature length scale $\lambda_{\text{R}}$, which is given by the inverse of the square root of the largest eigenvalue of the Riemann tensor of the background spacetime. In total, we have the following hierarchy of scales
\begin{equation}
r_+ \ll l \ll \lambda_{\text{R}}.
\end{equation}
For simplicity, we consider our black hole to be static and spherically symmetric, although the thought experiment we describe in the following readily generalizes to rotating black holes.

Our aim is to determine the black hole's entropy. We can only do so indirectly, by letting the black hole absorb some matter and computing the corresponding change in its entropy from the first law of black hole thermodynamics. At the distance $l$ from the black hole we prepare a small box filled with matter, which we lower into the black hole along a radial geodesic. For simplicity, we assume that this matter is perfectly randomized, i.e., that no work can be extracted from it. Then, the entire energy absorbed by the black hole corresponds to heat transfer. We can conveniently evaluate this heat, $Q$, before we start lowering the box. We obtain
\begin{equation}
Q=-\int_{\text{B}}T^{\mu\nu}u_{\mu}u_{\nu}\text{d}^{D-1}\text{B},
\end{equation}
where the integral is carried out over the box $\text{B}$ with the spatial volume element $\text{d}^{D-1}\text{B}$ ($D$ denotes the spacetime dimension), $T^{\mu\nu}$ denotes the energy-momentum tensor of the matter inside $\text{B}$, $u_{\mu}$ is the spacetime velocity of $\text{B}$. Since $\text{B}$ is small and we can compute the integral at distance $l$ from the black hole, $Q$ has the same value as it would have in a flat spacetime, up to negligible corrections.

The first law of black hole thermodynamics implies~\cite{Gao:2001,Wall:2015,Visser:2024}
\begin{equation}
\label{first law}
Q=T_{\text{H}}\Delta S_\text{W},
\end{equation}
where $T_{\text{H}}=\kappa/\left(2\pi\right)$\footnote{We work in Planck units, setting $G=\hbar=c=k_{\text{B}}=1$.}, with $\kappa$ being the black hole surface gravity, is the Hawking temperature at which the black hole radiates~\cite{Hawking:1975}. The quantity $\Delta S_\text{W}$ is then the change of the black hole's Wald entropy~\cite{Wald:1993,Wald:1994,Hollands:2024,Visser:2024}.

Black hole entropy is known to scale with the area $\mathcal{A}$ of the spatial cross-section of the black hole's horizon~\cite{Hollands:2024}. Both $\mathcal{A}$ and $\kappa$ (and, consequently, $T_{\text{H}}$) are measurable characteristics of the black hole. Since we also know the absorbed heat $Q$, we can in principle measure the value of the dimensionless ratio
\begin{equation}
\label{ratio}
\mathcal{S}\equiv \frac{Q}{T_{\text{H}}\Delta\mathcal{A}}=\frac{\Delta S_{\text{W}}}{\Delta\mathcal{A}},
\end{equation}
where the second equality follows from the first law~\eqref{first law}. In the limit of infinitesimal $\Delta\mathcal{A}$, $\mathcal{S}$ would have the interpretation of the areal density of entropy.

If the ratio $\mathcal{S}$ depends on the properties of the background spacetime, we can use its value to ``label'' spacetime points. Since $\mathcal{S}$ is a characteristic of a local test physical process, such labeling directly violates the strong equivalence principle.

There is one potential problem we need to address. Tidal forces caused by the background curvature generically deform the shape of the apparent horizon. The deformations may remain significant even in the limit of a very small black hole. Then, we need to check whether the background curvature also noticeably affects the horizon area, $\mathcal{A}$, and, consequently, the value of $\mathcal{S}$ in this way. Using the Riemann normal coordinate expansion around the metric of the static, spherically symmetric black hole~\cite{Brewin:2009}, we can compute the area in a straightforward way, obtaining
\begin{equation}
\mathcal{A}=r_+^{D-2}\Omega_{D-2}+O\left(r_+^{D}\right),
\end{equation}
where $\Omega_{D-2}$ denotes the area of unit, $\left(D-2\right)$-dimensional sphere and the $O\left(r_+^{D}\right)$ contributions depend on the background spacetime curvature. For sufficiently small $r_+$, only the first term is relevant and $\mathcal{A}$ is, with the necessary precision, independent of the background curvature. In other words, even relatively large tidal deformations of the horizon only lead to negligible changes of $\mathcal{A}$. The background curvature can only affect $\mathcal{S}$ through Wald entropy, whose form is directly determined by the gravitational dynamics.

In the following, we consider our thought experiment in several notable gravitational theories. In each case, we simply need to determine the corresponding Wald entropy and compute the dimensionless ratio $\mathcal{S}$ via equation~\eqref{ratio}. Any dependence of the resulting $\mathcal{S}$ on the background spacetime violates the strong equivalence principle.

\section*{General relativity}

To test our approach, we first consider the case of general relativity, for which the strong equivalence principle is known to be satisfied~\cite{Casola:2014,Casola:2015}. In this case, Wald entropy simply equals Bekenstein entropy~\cite{Bekenstein:1973}, which is proportional to the area of the black hole's apparent horizon~\cite{Hollands:2024,Visser:2024}. The change of this entropy in our thought experiment reads
\begin{equation}
\Delta S_{\text{W}}=\frac{\Delta\mathcal{A}}{4}.
\end{equation}
Hence, according to equation~\eqref{ratio}, $\mathcal{S}$ is a universal constant, $\mathcal{S}=1/4$. As expected, black hole thermodynamics in general relativity does not violate the strong equivalence principle\footnote{As an aside, the same result $\mathcal{S}=1/4$ also holds for Weyl transverse gravity~\cite{Carballo:2022}, for which we have recently shown the validity of the strong equivalence principle~\cite{equivalence}.}.

\section*{$f\left(R\right)$ gravity}

Next, we discuss the simplest class of modified gravitational Lagrangians that can be written as an arbitrary function of the scalar curvature, $f\left(R\right)$. The corresponding Wald entropy reads~\cite{Jacobson:1993}
\begin{equation}
S_{\text{W}}=\frac{1}{4}\int f'\left(R\right)\text{d}\mathcal{A},
\end{equation}
where the integration is carried out over a spatial cross-section of the apparent horizon. Since the black hole is very small, we can Taylor expand the integrand, keeping only the leading order in $r_+$. The change of Wald entropy then obeys
\begin{equation}
\Delta S_{\text{W}}=\frac{f'\left(R\right)}{4}\Delta\mathcal{A},
\end{equation}
where we approximate the scalar curvature by its value at some arbitrary point $P$ on the horizon. Thus, the ratio $\mathcal{S}$ equals
\begin{equation}
\mathcal{S}=\frac{f'\left(R\right)}{4}.
\end{equation}
Since $R$ is the background scalar curvature, $\mathcal{S}$ explicitly depends on the position in spacetime (with the exception of general relativity, for which $f'\left(R\right)=1$). Moreover, as $\mathcal{S}$ is independent of the properties of the test black hole, the dependence on the scalar curvature cannot be removed by taking the limit of a sufficiently small black hole. Then, we can distinguish the different spacetime points by measuring $\mathcal{S}$ in our thought experiment. This outcome directly violates the strong equivalence principle.

The violation of the strong equivalence can be traced to the presence of dimensionful coupling constants in the gravitational Lagrangian. Such constants lead to terms in Wald entropy which depend on the spacetime curvature and, at the same time, do not scale with the horizon radius $r_+$, so they appear for arbitrarily small black holes. For instance, if $f\left(R\right)=R+\alpha R^2$, with $\alpha$ being a coupling constant of length dimension $2$, the ratio $\mathcal{S}$ equals $\mathcal{S}=\left(1+2\alpha R\right)/4$. Consequently, any local, metric gravitational theory whose Lagrangian contains dimensionful couplings is incompatible with the strong equivalence principle.

\section*{Lanczos-Lovelock gravity}

Lanczos-Lovelock gravity represents the most general class of local, purely metric theories whose equations of motion are at most second order in the derivatives of the metric~\cite{Lanczos:1938,Lovelock:1971,Padmanabhan:2013}. Any Lagrangian $L_{\text{LL}}$ in this class can be written as a sum of terms of the form
\begin{equation}
L^{\left(k\right)}=\alpha_{k}\frac{\left(2k\right)!}{2^{k}}\,\delta_{[c_{1}}^{a_{1}}\delta_{d_{1}}^{b_{1}}\dots\delta_{c_{k}}^{a_{k}}\delta_{d_{k}]}^{b_{k}}R_{a_{1}b_{1}}^{\quad\;\, c_{1}d_{1}}\ldots R_{a_{k}b_{k}}^{\quad\;\, c_{k}d_{k}},
\end{equation}
where $k$ is a natural number, $\alpha_{k}$ a coupling constant with length dimensions $2\left(k-1\right)$ and the square bracket denotes total antisymmetrization of the indices. In $D=4$ spacetime dimensions, the only nontrivial Lanczos-Lovelock Lagrangian is that of general relativity. In higher dimensions, there exists other nontrivial Lanczos-Lovelock theories, such as Einstein-Gauss-Bonnet gravity.

With the exception of general relativity, Lanczos-Lovelock theories depend on dimensionful couplings. Then, it appears clear that their thermodynamics violates the strong equivalence principle by the same reasoning we discussed on the example of $f\left(R\right)$ gravity. Nevertheless, we have a good reason to treat them separately. Lanczos-Lovelock gravity is the only class of theories known to satisfy the gravitational weak equivalence principle, which requires that test particles with significant self-gravity (in particular, small black holes) move in vacuum independently of their properties~\cite{Casola:2014}. The gravitational weak equivalence principle together with the local Poincaré invariance of test physics then directly implies the strong equivalence principle. From this perspective, Lanczos-Lovelock gravity is a likely candidate for the most general class of theories compatible with the strong equivalence principle. However, an explicit check of this hypothesis has been so far lacking. With our thermodynamic approach, it turns out to be straightforward.

The change of Wald entropy in our thought experiment for Lanczos-Lovelock gravity equals
\begin{equation}
\Delta S_{\text{W}}=\frac{P^{tr}_{\;\:\;tr}}{2}\Delta\mathcal{A},
\end{equation}
where we introduced the tensor
\begin{equation}
P^{\mu\nu\rho\sigma}=\frac{\partial L_{\text{LL}}}{\partial R_{\mu\nu\rho\sigma}},
\end{equation}
and the component $P^{tr}_{\;\:\;tr}$ is written in the Boyer-Lindquist coordinates~\cite{Boyer:1967}. The corresponding $\mathcal{S}$ reads
\begin{equation}
\mathcal{S}=\frac{P^{tr}_{\;\:\;tr}}{2}.
\end{equation}
For any theory except general relativity, $P^{tr}_{\;\:\;tr}$ explicitly depends on the background Riemann tensor. Therefore, Lanczos-Lovelock gravity violates the strong equivalence principle.

To provide a concrete example, Einstein-Gauss-Bonnet gravity, the most general Lanczos-Lovelock Lagrangian in $D=5,6$, yields the following ratio $\mathcal{S}$
\begin{equation}
\mathcal{S}=\frac{1}{4}+\frac{1}{2}\alpha_2\left(R-4R_{t}^{\;r}+2R^{tr}_{\;\:\;tr}\right),
\end{equation}
which indeed depends on the background curvature.

Since the gravitational weak equivalence principle is satisfied, the only possible explanation of our result lies in the violation of the local Poincaré invariance of the gravitational test physics. Remarkably, there exists further support for this interpretation. It has been argued that test gravitational waves in Lanczos-Lovelock gravity do not always propagate at the speed of light~\cite{Izumi:2014,Papallo:2015,Benakli:2016,Camanho:2016}, which indeed suggests that their motion cannot be locally Poincaré invariant.

\section*{Scalar-tensor gravity}

We now extend our analysis to theories with nonminimal coupling between gravity and matter fields. While, for concreteness, we focus on two particular scalar-tensor theories, our findings readily generalize to arbitrary scalar-tensor Lagrangians.

We start with Brans-Dicke gravity, historically the first scalar-tensor Lagrangian to be proposed~\cite{Brans:1961}
\begin{equation}
L_{\text{BD}}=\phi R+\frac{\omega\left(\phi\right)}{\phi}\nabla_{\lambda}\phi\nabla^{\lambda}\phi+V\left(\phi\right),
\end{equation}
where $\omega\left(\phi\right)$, $V\left(\phi\right)$ are arbitrary functions of the scalar field. The corresponding change in Wald entropy equals
\begin{equation}
\Delta S_{\text{W}}=\frac{\phi}{4}\Delta\mathcal{A},
\end{equation}
Consequently, the ratio $\mathcal{S}$ is
\begin{equation}
\mathcal{S}=\frac{\phi}{4},
\end{equation}
which explicitly depends on the background value of the scalar field $\phi$. Therefore, Brans-Dicke theory is not compatible with the strong equivalence principle. The same conclusion applies to any Lagrangian which contains the term $\phi R$.

The second example of a scalar-tensor Lagrangian we consider reads
\begin{equation}
L_{5}=R+2\gamma G^{\lambda\rho}\nabla_{\lambda}\phi\nabla_{\rho}\phi,
\end{equation}
where $G^{\lambda\rho}$ denotes the Einstein tensor. This theory is notable for introducing no modifications to the Wald entropy of stationary black holes~\cite{Feng:2016}. Then, it holds
\begin{equation}
\Delta S_{\text{W}}=\frac{\Delta\mathcal{A}}{4},
\end{equation}
and we recover the general relativistic result $\mathcal{S}=1/4$.

Our thought experiment detects no violation of the strong equivalence principle for this theory. However, it is known to both violate the gravitational weak equivalence principle~\cite{Casola:2014} and to imply modified propagation speed of test gravitational waves~\cite{Kobayashi:2012,Kobayashi:2014}. Thence, the Lagrangian $L_{5}$ cannot satisfy the strong equivalence principle. It serves to show that the independence of $\mathcal{S}$ on the position in spacetime is necessary but by no means sufficient condition for the validity of the strong equivalence principle.

\section*{Area-dependent entropies}

To conclude, we discuss a class of entropy functions that do not follow from any local gravitational Lagrangian. In many approaches, black hole entropy can be written as some function $\chi$ of the horizon area, i.e.,
\begin{equation}
\label{mod ent}
S=\chi\left(\mathcal{A}\right).
\end{equation}
This general expression covers, e.g. the logarithmic corrections expected to arise due to quantum gravitational effects~\cite{Solodukhin:1995,Kaul:2000,Medved:2005,Banerjee:2011,Faulkner:2013,Xiao:2021}, the Barrow prescription which assumes a fractal structure of the horizon~\cite{Barrow:2020}, and the non-extensive Tsallis entropy~\cite{Tsallis:1988,Tsallis:2022}.

In this case, we have for $\mathcal{S}$
\begin{equation}
\mathcal{S}=\frac{\Delta\chi\left(\mathcal{A}\right)}{4\Delta\mathcal{A}}.
\end{equation}
On the one hand, $\mathcal{S}$ depends explicitly on $\mathcal{A}$, unless $\chi\left(\mathcal{A}\right)=\mathcal{A}$. On the other hand, it is completely determined by the intrinsic properties of the test black hole and independent of the background spacetime. Consequently, we cannot learn about the position in the spacetime by measuring $\mathcal{S}$ for entropy prescriptions of the form~\eqref{mod ent}. Our thought experiment then does not uncover any violation of the strong equivalence principle. The same conclusion applies to any entropy which depends solely on the intrinsic properties of the test black hole.

\subsection*{Conclusions}

The examples we have discussed, of course, do not exhaust the power of our method. In the future, our thought experiment can be applied in the context of, e.g. semiclassical gravity, AdS/CFT correspondence, nonlocal theories, and other situations in which the status of the strong equivalence principle remains unclear. Given that the only necessary input is the expression for black hole entropy, it becomes relatively easy to look for violations of the strong equivalence principle in a wide class of models.

We already obtained a significant new result, the violation of the strong equivalence principle in Lanczos-Lovelock gravity. It has been known that the compatibility with the gravitational weak equivalence principle selects Lanczos-Lovelock gravity among all the metric theories of gravity. We show that the strong equivalence principle restricts our freedom even more, leaving general relativity as the only possible local, diffeomorphism-invariant description of gravity that does not violate it. This example further demonstrates how tests for the different equivalence principles can be used together to yield novel insights about the theory in question. For Lanczos-Lovelock gravity, it shows that the test gravitational physics must violate the local Poincaré invariance~\cite{Casola:2015}, which indeed appears to be the case~\cite{Izumi:2014,Papallo:2015,Benakli:2016,Camanho:2016}.

So far, we have discussed our criterion as a thought experiment with some theoretical consequences. Can it ever be realized as an actual measurement? Astonishingly, there exists a precedent for an observational study of black hole thermodynamics. The area increase theorem and, indirectly, the generalized second law of thermodynamics in a black hole merger has been verified by analyzing the LIGO data~\cite{Cabero:2018}. Thus, to conclude, let us imagine an observational check of the strong equivalence principle.

In practice, the test black hole can be fairly large. Indeed, the observed astrophysical black holes are very well described by the Kerr metric, indicating that they are not significantly affected by the background gravitational field. Of course, Kerr black holes rotate and, thence, are not static and spherically symmetric as in our model. However, accounting for the effect of rotation in our criterion is simple. Then, observing the fall of some matter into the black hole, we might be able to obtain the necessary input for evaluating our criterion. The area of the horizon of a Kerr black hole can be computed from observational data~\cite{Cabero:2018}, and the same applies to surface gravity. The mass-energy of the in-falling matter can also in principle be independently determined, giving us all the necessary information. An actual observation of this type would be indeed very difficult. Nevertheless, given its remarkable simplicity, our criterion for the strong equivalence principle might possibly have relevance even beyond its purely theoretical application.

\section*{Acknowledgements}

AA-S is funded by the Deutsche Forschungsgemeinschaft (DFG, German Research Foundation) — Project ID 51673086. ML is supported by the DIAS Post-Doctoral Scholarship in Theoretical Physics 2024. AA-S and LJG acknowledge support through Grant No. PID2023-149018NB-C44 (funded by MCIN/AEI/10.13039/501100011033). LJG also acknowledges the support of the Natural Sciences and Engineering Research Council of Canada (NSERC).

\bibliography{bibliography}

\end{document}